# Intereses convergentes: los años previos a la visita de Einstein a la Argentina en 1925

Alejandro Gangui

Universidad de Buenos Aires, Facultad de Ciencias Exactas y Naturales, Argentina. CONICET – Universidad de Buenos Aires, Instituto de Astronomía y Física del Espacio (IAFE), Argentina

Analizamos en cierto detalle los años previos a la visita de Albert Einstein a la Argentina, desde que su figura se tornó célebre luego de la publicación de los resultados del eclipse de Sol del año 1919, hasta su arribo al puerto de Buenos Aires en marzo de 1925. Veremos que este evento mayor en la vida cultural del país no surgió del vacío, sino que fue el fruto de variados intereses y de singulares circunstancias internacionales y locales, donde la situación social y política que rodeaba al eminente científico y el impulso de individuos e instituciones de Buenos Aires jugaron un rol determinante y que a la larga convergieron en la recepción "exitosa" de este personaje tan fuera de lo común en un país de la periferia científica del primer cuarto del siglo XX. Reseñamos también las referencias escritas y los conocimientos que circulaban en el ambiente local sobre temas conexos con la teoría de la relatividad en años previos a su arribo al país.

**Palabras clave:** historia de la ciencia, teoría de la relatividad, Argentina, siglo XX.

We analyze in some detail the years leading up to Albert Einstein's visit to Argentina, from the moment he rose to fame after the publication of the results of the 1919 solar eclipse, until his arrival at the port of Buenos Aires in March 1925. We will see that this major event in the country's cultural life did not arise from a vacuum but was the product of diverse interests and unique international and local circumstances. The social and political situation surrounding the eminent scientist and the impetus provided by individuals and institutions in Buenos Aires played a decisive role, ultimately converging in the "successful" reception of this extraordinary figure in a country on the scientific periphery of the first quarter of the 20th century. We also review the written references and knowledge circulating in the local milieu on topics related to the theory of relativity in the years prior to his arrival in the country.

**Keywords:** history of science, theory of relativity, Argentina, 20th century.

I. Introducción

El largo viaje de Albert Einstein a Sudamérica, en el que pasó todo un mes en la Argentina, fue una empresa que comenzó a concebirse con varios años de antelación. El indicio más certero de cómo surgió la idea inicial de hacer que el padre de la relatividad se desplazara por vez primera del otro lado del ecuador no recae en algún miembro de la comunidad científica o política de nuestro país sino, quizás sorprendentemente, en un poeta. Leopoldo Lugones era en esos años uno de los principales escritores y poetas locales, de gran influencia en el ambiente de la prensa y de la cultura porteña, y asiduo concurrente de tertulias culturales donde se reunían los principales exponentes de las artes, las letras y las ciencias, y él mismo amigo de varios académicos y científicos. El contacto con las novedosas ideas que iban surgiendo en varios ámbitos de la ciencia hicieron de Lugones un lector asiduo de los últimos descubrimientos y teorías provenientes de Europa. Y, entre los avances que más llamaron su atención, sin duda, la relatividad de Einstein ocupó un lugar privilegiado.

En julio de 1921, durante su visita a París como representante argentino en el Comité Internacional de Cooperación Intelectual de la Liga de las Naciones, Lugones hizo un llamado a sus colegas franceses para que le indicasen el nombre de algún experto que pudiese viajar a la Argentina y brindar un curso académico sobre temas avanzados de relatividad. Es bueno recordar que en aquel momento ya existía una cierta masa crítica de científicos, ya sea alemanes o que habían sido entrenados en Alemania, que investigaban y hacían docencia en los primeros centros de física de nuestro país, principalmente en la ciudad de La Plata, pero también en Córdoba y en Buenos Aires. Por ejemplo, el físico Richard Gans dirigía el Instituto de Física de La Plata, el fisiólogo Georg Friedrich Nicolai daba clases en Córdoba y el matemático español, con estudios superiores en Alemania, Julio Rey Pastor, impulsaba el desarrollo de la matemática avanzada en Buenos Aires. Ante el pedido de Lugones, es natural que las autoridades alemanas reaccionaran con premura. Tanto el Ministerio de Asuntos Extranjeros alemán como el Ministerio Prusiano de Educación contactaron a Einstein para saber si él mismo podría viajar y hacer un tour de conferencias en respuesta al pedido de Lugones. La correspondencia del científico con los representantes de estos ministerios indica que Einstein no tenía posibilidad de viajar a Sudamérica en el corto plazo y por eso sugería que su lugar fuese tomado por Jakob Laub, un joven físico y colega suyo que conocía muy bien los diversos temas relativistas. Estas tratativas finalmente no se concretaron y debió pasar casi un año hasta que Lugones volviera a la carga con su idea inicial.

Hacia mediados de 1922, el asesinato del ministro de relaciones exteriores de la República de Weimar y la noticia de las amenazas sufridas por Einstein llevó a Lugones a volver sobre la cuestión de divulgar la relatividad en nuestro país, pero ahora con la motivación adicional de "rescatar" al científico y alejarlo de Alemania al menos hasta que la situación social que lo rodeaba se pacificase. Como veremos en más detalle en las páginas que siguen, Lugones propuso entonces traer a Einstein a la Argentina, crear una cátedra universitaria nueva especial para él y brindarle las condiciones adecuadas como para que pudiera trabajar en sus investigaciones y enseñar su teoría en paz. Para ese momento tanto Einstein como la verificación de su teoría general de la relatividad atraían la atención de los diarios de mayor circulación del país y el interés científico se había materializado en al menos algunos libros y en varias conferencias y artículos para un público amplio. Sin embargo, estos trabajos relativistas no eran una novedad en los círculos más restringidos de la academia, ya que mucho antes de que Einstein fuera visto como una celebridad científica, los trabajos sobre temas que hoy podríamos llamar precursores de la relatividad ya habían hecho su aparición, aunque modesta, en las revistas científicas locales del momento (Galles, 1982; Ortiz, 1995).

## II. La relatividad en la Argentina

En la primera década del siglo XX comenzaron a aparecer algunos estudios sobre la materia y la radiación, y sobre la nueva teoría de la relatividad (restringida) de Einstein. Esta teoría de 1905 ofrecía una visión unificada de la materia y la energía e introducía concepciones nuevas sobre la estructura del espacio físico. Esos estudios producidos en el ambiente local tenían un carácter informativo serio y se difundieron en algunos de los medios de prestigio científico del momento, por ejemplo, en revistas universitarias, en revistas de ingeniería y en los *Anales* de la Sociedad Científica Argentina. Los autores de estos artículos eran tanto argentinos como extranjeros y la mayoría de ellos eran miembros acreditados del sector académico [Fig. 1].

**SIN HACER REFERENCIA AL TRABAJO DE EINSTEIN:**

- **de Lepiney, P. 1906-8**, Los electrones y las radiaciones, *Rev. Técnica* (Buenos Airess, tres artículos s/ la dinámica del electrón de Max Abraham de 1902).
- **de Lepiney, P. 1907**, La dinámica sin el segundo principio de Newton, *Anales UBA* (usa m=m(v)).
- **Broggi, U. 1909**, Sobre el principio electrodinámico de relatividad y sobre la idea de tiempo, *Rev. Politécnica* (futura revista del CEI).

**CON REFERENCIA AL TRABAJO DE EINSTEIN:**

- **Volterra, V. 1910**, Espacio, tiempo i masa (conferencia), *Anales SCA*. (primera mención al trabajo de Einstein)
- **Laub, J. 1911**, "Curso especial de tipo teórico" en La Plata (s/ mecánica avanzada, incluye relatividad, Einstein, Minkowski, Planck; cf. carta de Emil Bose a Laub, 1911, es el primer curso regular universitario s/ la relatividad en el Nuevo Mundo).
- **Laub, J. 1912**, Noticia sobre "los efectos ópticos en medios en movimiento", *Anales SCA*.
- **Duclout, J. 1914**, en *Novedades científicas*, Conferencia en la U. de Tucumán (sobre concepto de masa. Menciona a Lorentz, Minkowski, Einstein).
- **Meyer, C. 1910-15**, "Cursos optativos física-matemática", UBA. Curso 1914: principio de relatividad, experiencias de Kaufmann (1901-03 s/ dependencia de la velocidad de la masa electromagnética del electrón).
- **Laub, J. 1915**, Propagación de la luz en los cuerpos en movimiento (folleto INPS), *Anales SCA*.
- **Laub, J. 1916**, Los teoremas energéticos y los límites de su validez, *Rev. de Filosofía*.
- **Laub, J. 1919**, ¿Qué son espacio y tiempo?, *Rev. de Filosofía*.

**Figura 1.** Primeras publicaciones, conferencias y cursos en la Argentina relacionados con temas de la relatividad, antes de que los trabajos fundacionales fueran conocidos y años más tarde, con mención explícita a Einstein.

Podemos mencionar algunos de los investigadores extranjeros, visitantes temporarios o ya afincados en nuestro país, que contribuyeron a elevar el nivel de la discusión académica en el área de la física teórica que estamos considerando. Entre los trabajos publicados en nuestro medio sobre temas relacionados con la relatividad se destaca el estudio de 1909 del matemático italiano Ugo Broggi, discípulo de David Hilbert. Este es un análisis de la materia, la radiación y el tiempo con un contenido científico preciso. Cuando apareció este trabajo Broggi era profesor contratado en la Universidad de La Plata. Un año más tarde el conocido fisicomatemático y senador italiano Vito Volterra visitó la Argentina como delegado oficial de Italia a las celebraciones del primer Centenario de nuestra independencia. Una vez en Buenos Aires Volterra ofreció una conferencia ante la Sociedad Científica Argentina (SCA) y en ella se ocupó del espacio-tiempo y la materia e hizo referencia explícita al trabajo de Einstein (Volterra, 1910). Entre 1910 y 1915 el fisicomatemático francés Camilo Meyer, amigo y compañero del secundario en la ciudad de Nancy del célebre matemático Henri Poincaré, dictó una serie de cursos libres sobre física-matemática en la Universidad de Buenos Aires (UBA) y en el último de ellos introdujo por primera vez en América Latina una exposición detallada de la teoría cuántica (Meyer, 1915). En ese curso hizo referencias explícitas a los primeros trabajos importantes de Einstein, pero, por lo específico de su tema, no a aquellos que se relacionaban directamente con la teoría de la relatividad.

Por otra parte, a fines de la primera década de 1900 el prestigioso astrónomo estadounidense Charles Dillon Perrine, del Observatorio Lick, en California, fue contratado como director del Observatorio de Córdoba, en ese entonces llamado el Observatorio Nacional Argentino, que había sido creado a comienzos de la década de 1870 y reactivado hacia 1910 dentro de un nuevo estímulo al desarrollo de las ciencias experimentales. Perrine tuvo a su cargo la renovación de los talleres de óptica y las facilidades de observación de esa institución y en 1912, respondiendo a un pedido de colegas suyos en Alemania asociados con Einstein, principalmente Erwin Freundlich, inició una serie de observaciones durante eclipses totales de Sol tendientes a detectar una posible deflexión de la luz de las estrellas lejanas al pasar en cercanías del limbo solar, predicción crítica de la teoría de la relatividad general de Einstein. Esas observaciones comenzaron en Brasil en 1912 (Paolantonio, 2019), continuaron luego en la península de Crimea, a orillas del mar Negro, en 1914, y en Venezuela en 1916, pero, debido a condiciones meteorológicas desfavorables, Perrine y sus colaboradores no lograron atacar el problema de la alteración de la luz estelar propuesto por la relatividad general (Gangui y Ortiz, 2009). Las observaciones concluyentes buscadas por los astrónomos para verificar dicha teoría solo llegarían en el año 1919. Lamentablemente, la Argentina no se hizo presente en esa ocasión tan favorable.

Además de los científicos que acabamos de listar, a comienzos de la década de 1910 la Universidad Nacional de La Plata contrató al ya mencionado físico Jakob Laub como profesor de física. Antes de su llegada a la Argentina Laub había tenido un contacto personal con Einstein (Seelig, 1954) y había colaborado con él en publicaciones sobre temas relativistas (Pyenson, 1985). A partir de 1912 Laub comenzó a publicar una serie de trabajos sobre la teoría de la relatividad en los *Anales de la SCA*. En la Argentina, las suyas fueron las contribuciones de un científico que conocía las ideas de Einstein y su teoría en detalle. De ahí que, como vimos más atrás, el mismo Einstein propusiera su nombre para reemplazarlo en el ciclo de conferencias que pedía Lugones en 1921. Aunque las publicaciones de Laub sobre la relatividad comenzaron en revistas científicas de carácter técnico [Fig. 1], a medida que la década avanzaba y el interés público por la relatividad crecía en muchos círculos de intelectuales, se movieron hacia calificados medios de difusión general. Por ejemplo, Laub comenzó a publicar en la *Revista de Filosofía* que era dirigida por el prestigioso médico y filósofo positivista José Ingenieros. Esta revista había sido creada en 1915 cuando, durante la Primera Guerra Mundial, las revistas europeas dejaron de llegar al río de la Plata; claramente, estaba orientada hacia un público con un interés cultural mucho más amplio que el de la física teórica.

**III. La relatividad en la Argentina en años posteriores al eclipse de 1919**

El 29 de mayo de 1919 comenzó un vertiginoso proceso que en muy breve tiempo llevó a un punto de inflexión en la figura pública de Einstein. Ese día se logró observar un eclipse total de Sol que permitió verificar una de las predicciones más importantes de la teoría de la relatividad general: la deflexión de la luz de las estrellas lejanas en el espacio curvo originado por un objeto de gran masa, en este caso, el Sol (Eddington y Crommelin, 1920), todo esto al poco tiempo del final de la Primera Guerra mundial (Stanley, 2003). A partir de ese año, que en adelante fue conocido como el año del "eclipse de Einstein" (Gangui y Ortiz, 2009), comenzaron a aparecer, con frecuencia, noticias y artículos sobre diversos aspectos de la teoría de la relatividad y sobre la actualidad de las discusiones científicas en Europa sobre temas conexos. Como vimos, la Argentina no fue ajena a ese interés en las publicaciones científicas; tampoco lo fue en la repercusión que la relatividad tuvo ante el público educado y la prensa.

Como señalamos páginas atrás, es indudable que Lugones fue un responsable principal de la presencia de estos temas en los medios locales. El interés del poeta por la ciencia, y por la relatividad de Einstein en particular, está muy bien documentado. Además, Lugones ejercía una influencia intelectual directa y destacada sobre los editores de varios de los cotidianos de gran circulación en el país, entre los que él mismo se contaba; muchos de ellos eran también poetas y escritores renombrados de esa época.

Aunque las lecturas científicas de Lugones eran variadas y frecuentes, ello no lo convertía en un experto, y menos aún en temas de física teórica avanzada. Sin embargo, era un intelectual muy respetado y su juicio sobre casi cualquier tema, incluso en temas científicos, tenía un peso considerable (Asúa y Hurtado de Mendoza, 2006). En 1920 ofreció una conferencia sobre las implicancias de la teoría de la relatividad de Einstein frente al público del Centro de Estudiantes de Ingeniería de la Facultad de Ciencias Exactas, Físicas y Naturales de la UBA, conocida entonces como "Facultad de Ingeniería". En base a esa conferencia, el año siguiente Lugones publicó un pequeño libro titulado *El tamaño del espacio* (Lugones, 1921). Y aunque esta obra no estaba libre de interpretaciones erróneas, de citas algo extemporáneas o de párrafos oscuros y aun esotéricos, su publicación sirvió para colocar a la relatividad en una posición favorable frente a la audiencia que se interesaba por los temas científicos en el Buenos Aires de la época (Hurtado de Mendoza y Asúa, 2005). Claramente, que un intelectual del prestigio de Lugones se interesase por una teoría en principio tan alejada de la cotidianeidad, era mucho más importante para la recepción de esa teoría que la precisión en la descripción de la física que el poeta pudiera lograr. A este trabajo siguieron muchos otros escritos de autores locales y también traducciones de trabajos extranjeros sobre temas relativistas [Fig. 2].

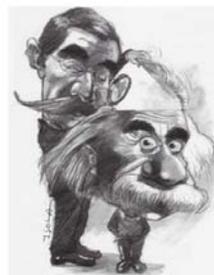

**Figura 2.** Trabajos escritos, conferencias y otros materiales publicados sobre temas de la teoría de la relatividad aparecidos en años posteriores al eclipse de Einstein.

A partir de ese año, los principales diarios de Buenos Aires incluyeron de forma habitual noticias acerca de los viajes y de los múltiples tributos que el sabio alemán recibía en diferentes países del mundo. Un mes más tarde, y en la misma tribuna que lo hiciera Lugones, también invitado por el Centro de Estudiantes de Ingeniería, el ingeniero y físico teórico Jorge Duclout leyó una ponencia sobre materia-energía y relatividad que luego fue publicada (Duclout, 1920). Duclout, amigo cercano de Lugones a quien le dedicó este trabajo, era un ingeniero alsaciano que se había afincado en el país en la década de 1880. Conocedor de herramientas matemáticas sofisticadas, tenía además una cultura sólida en ciencias físicas. En los años previos al cambio de siglo organizó seminarios de matemática avanzada en la SCA. Duclout se había graduado en el prestigioso Politécnico de Zürich (hoy conocido como ETH), al igual que más tarde lo hizo Einstein, y tuvo, como veremos en las próximas páginas, un rol importante en la materialización de la visita del sabio alemán (Gangui y Ortiz, 2008).

A esta lista de publicaciones post eclipse podemos agregar, además, las conferencias específicamente sobre la relatividad que dictó el físico español Blas Cabrera durante su visita a la Argentina en 1920. También, el trabajo de otro español, el astrónomo Jesuita José Ubach (profesor en el Colegio del Salvador de Buenos Aires donde mantenía un observatorio) sobre los resultados de la expedición astronómica de 1919 (Ubach, 1920). Estas fueron contribuciones detalladas, serias y orientadas a los intereses de un público culto. Por otra parte, el ya mencionado matemático puro Julio Rey Pastor, también de origen español, pero en ese momento ya residente en la Argentina, contribuyó con trabajos relacionados con la relatividad, principalmente en su relación con la geometría (Ortiz, 2011). Rey Pastor había conocido a Einstein en Berlín y, más cerca de la visita de este último a la Argentina, publicó notas divulgativas en tópicos relacionados con la relatividad en diarios locales, particularmente en el diario argentino *La Nación*. El temario de algunos de los cursos universitarios más avanzados de las ciencias exactas también tuvo un impacto positivo. Nuevamente, Rey Pastor incluyó temas propios de la matemática de la relatividad, como ser la geometría diferencial moderna, en sus seminarios de matemática avanzada en 1923 y 1924, poco antes de la visita de Einstein a la Argentina (Rey Pastor, 1989) [Fig. 3].

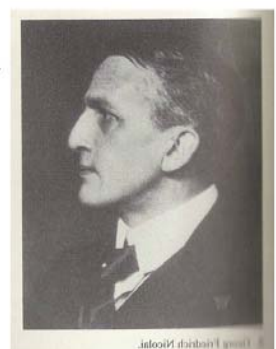

**PUBLICACIONES LOCALES POST ECLIPSE (II)**

- **1922**: traducción *paper* 1916 s/ RG, pedido de CEI a Einstein.
- **Butty, E. 1922-4**, en *Revista del CEI* (serie de artículos).
- **1922-3**: Conferencias en la SCA s/ la relatividad (Broggi, Rey Pastor, Gans, Collo, Isnardi, Aguilar, Damianovich).
- **Gilli, J. 1923**, La teoría de la relatividad, revista *Nosotros* (2 artículos, por sugerencia de J. Ingenieros).
- **Gilli, J. ~1925**, Einstein explicado, Librería García Santos (libro de 278 pp. con figuras).
- **Korn, A. 1923**, Kant, Einstein y Morente, *Rev. de Filosofía*.
- **Butty, E. 1924**, Introd. filosófica a las teorías de la relatividad.
- **Dellepiane, L. 1924**, Preocupaciones einsteinianas (Conf. en la Facultad de Ciencias, publicada en *Revista del CEI*).
- **Rey Pastor <1925**, varios artículos s/ la relación relatividad-geometría en el diario La Nación.
- **1925**: el film se habría pasado en una sala porteña, La Nación 28/03 (film no recuperado).
- **Nicolai, G.F. 1925**, Sentido filosófico de la teoría de la relatividad, *Revista de Filosofía*.

Georg F. Nicolai

**Figura 3.** Más ejemplos de artículos y de otros materiales publicados en los medios locales sobre temas relativistas en años posteriores al eclipse de 1919 y previos a la visita de Einstein al país.

Para finalizar esta enumeración incompleta de trabajos, notemos que en el par de años previo a la visita de Einstein a nuestro país apareció un interesante estudio teórico sobre la relatividad, las teorías restringida y general, escrito por dos jóvenes físicos argentinos, José B. Collo y Teófilo Isnardi que, con anterioridad, habían hecho estudios avanzados de física en Alemania (Collo, 1923; Isnardi, 1923). Debido a la calidad de su factura este trabajo es paradigmático, dentro de las ciencias físicas, de un nuevo e importante giro intelectual hacia las ciencias teóricas y que se operaba en la Argentina de esos años. El trabajo de Collo e Isnardi incluía una tercera parte, escrita por el astrónomo argentino Félix Aguilar (Aguilar, 1924), entrenado también en Europa, donde discutió las implicaciones astronómicas de la relatividad (Asúa y Hurtado de Mendoza, 2006; Gangui y Ortiz, 2011). Notemos también que el interés del público por los temas que habían hecho célebre a Einstein tuvo un singular paralelo entre los estudiantes universitarios de ese momento. Nuevamente fueron las autoridades del Centro de Estudiantes de Ingeniería las que, aparentemente por sugerencia de Rey Pastor, escribieron a Einstein el 5 de abril de 1922 pidiéndole permiso para traducir y publicar en español el célebre trabajo técnico sobre la relatividad general *Die Grundlage der allgemeinen Relativitätstheorie*, originalmente aparecido en la revista *Annalen der Physik* en 1916. El argumento esgrimido para tal acción era el gran interés suscitado por esos tópicos en los círculos universitarios locales. Ante este inusual pedido, Einstein respondió por carta el 31 de mayo de ese mismo año aceptando la propuesta. El científico sugería, además, que la publicación no fuera en una revista sino en forma de libro, y que le reservaran para él el 20 por ciento de las ganancias en calidad de derechos de autor.

**IV. Antecedentes de la visita de Einstein a la Argentina**

Si volvemos ahora unos pocos años hacia el pasado veremos que el contexto social que rodeaba a Einstein en Berlín en ese período estaba lejos de ser ideal como para que el científico pudiera dar continuidad a sus investigaciones. Como sabemos, para el año 1922 la situación social en Alemania había comenzado a degradarse. Los periódicos de la Argentina reiteraban sus noticias sobre la inestabilidad política alemana y, hacia el final de junio de ese año, anunciaron el asesinato de Walther Rathenau, ministro de relaciones exteriores de la República de Weimar. El diario *La Nación* también publicó la noticia de que la vida de Einstein corría peligro, asunto que fue actualizado en los meses sucesivos. En efecto, en carta a Max Planck del 6 de julio de 1922, el mismo Einstein menciona que había recibido información sobre rumores acerca de que su vida peligraba (Einstein, 2012: Documento 266) [Fig. 4]. De hecho, hacia fines de 1923 (entre el 7 noviembre y la navidad de ese año), Einstein debió abandonar su país y viajar temporariamente a Holanda por las amenazas que había recibido, como veremos más abajo.

### 266. To Max Planck

Kiel, 6 July 1922

Dear Colleague

This letter is not easy for me to write; but it really does have to be done. I must inform you that I cannot deliver the talk I promised for the Scientists' Convention, despite my earlier definite commitment.[1] For I have been warned by some thoroughly reliable persons (many of them, independently) against staying in Berlin[2] at present and ⟨generally⟩ particularly against making any kind of public appearances in Germany.[3] For, I am supposedly among the group of persons being targeted by nationalist assassins. I have no secure proof, of course; but the prevailing situation now makes it appear thoroughly credible.[4] If it had been an action of substantial professional importance, I would not have let myself be swayed by such motives, but a merely formal act is involved that someone (e.g., Laue)[5] could easily perform in my place.[6] The whole difficulty arises from the fact that newspapers mentioned my name too often and thereby mobilized the riffraff against me. So there is no helping it besides patience and—leaving town. I ask you one thing: Please take this little incident with humor, as I myself do.

With amicable greetings, yours,

A. Einstein.

**Figura 4.** Einstein en peligro. Traducción al inglés del texto de la carta que Einstein envía a Max Planck el 6 de julio de 1922 en la que menciona los rumores acerca de que su vida corría peligro (Einstein, 2012: Documento 266).

Ante esta situación, y como adelantamos brevemente en la Introducción, Lugones volvió sobre su idea inicial de traer al padre de la relatividad a nuestras tierras. Así, el 9 de agosto de 1922, el poeta publicó en *La Nación* un artículo donde sugería que el sabio fuese invitado a la Argentina, y que sus admiradores locales se encargasen de contribuir con fondos para cubrir los gastos y crear una cátedra universitaria independiente especial para él (Lugones, 1922). Poco tiempo después de esta nota, dos agrupaciones estudiantiles de Buenos Aires, las del Centro de Estudiantes de Ingeniería y el análogo del Instituto Nacional del Profesorado Secundario, anunciaron públicamente su apoyo a la idea de Lugones de ofrecer a Einstein un lugar seguro donde vivir y trabajar hasta tanto la situación en Alemania fuera más estable. Dos científicos de renombre ya mencionados -Rey Pastor, consejero de la primera asociación, y Laub, en ese entonces jefe del departamento de física de la segunda institución- habrían dado su apoyo a estas medidas. Sin duda alguna, el ingeniero Duclout, como amigo personal de Lugones, también fue parte de este movimiento de opinión. Recordemos que el primero había dedicado su trabajo (Duclout, 1920) al poeta, quien, a su vez, dedicó su obra *El tamaño del espacio* a su amigo ingeniero.

Menos de dos semanas más tarde, el 22 de agosto de ese año, Duclout presentó ante el Consejo Directivo de la Facultad de Ciencias Exactas, Físicas y Naturales de la UBA la propuesta de otorgar a Einstein el título de doctor *honoris causa* en ciencias fisicomatemáticas por sus trabajos relativistas. Avalada por unanimidad, en solo unos pocos días su propuesta fue aceptada también por el Consejo Superior de la UBA. El profesor de literatura española de la Facultad de Filosofía y Letras de la misma casa de altos estudios, Mauricio Nirenstein, quien menos de tres años más tarde jugó un rol relevante durante la visita del sabio alemán (Gangui y Ortiz, 2008), era en ese momento secretario de la UBA y a la vez un miembro prominente de la Asociación Hebraica Argentina. Como veremos, esta conexión fue fundamental para la materialización de la visita de Einstein pocos años más tarde.

Pasados poco más de dos meses, el 30 de octubre de 1922, y secundado por otros académicos, fue nuevamente Duclout quién presentó ante el Consejo Superior de la UBA una propuesta formal para que Einstein fuese invitado a dictar una serie de conferencias sobre la teoría de la relatividad. La ocasión no podía haber sido mejor elegida, pues menos de dos semanas más tarde los matutinos porteños trajeron el anuncio de que se había otorgado a Einstein el premio Nobel de física correspondiente al año 1921, que había quedado vacante del año anterior. Como es sabido, Einstein no recibió ese premio por sus trabajos sobre la relatividad, que no era universalmente aceptada aun después de las pruebas astronómicas producto de la observación del eclipse de 1919, sino por su explicación del llamado efecto fotoeléctrico.

Esta era una época de viajes frecuentes para Einstein. Había sido invitado a visitar tanto Inglaterra como los Estados Unidos de Norteamérica en 1921, Francia y Japón en 1922 (fue en el barco donde recibió la noticia del comité Nobel), y Palestina y España durante el año siguiente. Como es de imaginar, en esos años estos eran viajes por mar que duraban largos meses. A partir de su correspondencia privada del 11 de abril de 1923, sabemos también que Einstein rechazó una oferta de Profesor en la Universidad de Columbia, Estados Unidos, que colegas que había conocido en su viaje de 1921 le habían ofrecido; oferta que no estaba desconectada de la cuestión de su seguridad personal en medio de la inestable situación de Berlín en esos años que mencionamos antes, o al menos de la incierta imagen de Alemania que se veía desde América. Por su parte, la prensa argentina siguió, con detalle y con un marcado interés, todos sus movimientos.

Llegado el mes de noviembre de 1923 la situación social en Berlín se volvió insostenible para Einstein, quien decidió abandonar de apuro su ciudad y refugiarse temporariamente en Leiden, Holanda, en donde trabajaba su colega el físico Paul Ehrenfest, hasta que se calmaran las aguas. Se sabe que hubo varios factores que podrían haber motivado a Einstein a abandonar su ciudad y su familia (con su esposa Elsa, sin embargo, mantuvo correspondencia). Estos factores fueron los violentos disturbios por los alimentos, que ya escaseaban en Berlín y en varias ciudades, y ataques físicos a la población judía, rumores de golpe de estado por parte de movimientos de extrema derecha y, nuevamente, el temor por las amenazas hacia su persona. Como ejemplo, sabemos que del 3 al 5 de noviembre el precio del pan se disparó en un factor 6; el 5 y el 6 de ese mes, muchos locales fueron saqueados y judíos atacados en la calle y en sus hogares; simultáneamente, corrían rumores de que extremistas de derecha avanzarían sobre Berlín (conforme la carta del 8 de noviembre de 1923 que Einstein envía a su esposa Elsa desde Leiden y las notas en el Documento 141 de Einstein, 2015). En esos tiempos convulsionados, Einstein recibió dos ofertas más de trabajo en el extranjero, una de ellas de parte de Albert A. Michelson quien le ofrecía un puesto de Profesor en el departamento de física de la Universidad de Chicago y, en carta del 5 de diciembre de 1923, le consultaba muy concretamente sobre el monto del salario que Einstein consideraba que debería percibir (Documento 169 de Einstein, 2015).

A pesar de esta tan delicada situación social en Alemania, y en particular en la vida personal de Einstein, en Buenos Aires los esfuerzos por acercar al sabio a sus orillas continuaban a paso firme. Un año más tarde de la última presentación de Duclout ante el Consejo Superior de la UBA, el 21 de diciembre de 1923, el mismo Consejo se reunió en sesión extraordinaria y, entre sus temas de discusión, figuraba nuevamente la visita de Einstein. En esa instancia, se tomó en consideración una nota de la Asociación Hebraica que señalaba que ya se habían establecido contactos con el padre de la relatividad (con relación a esto, ver la Fig. 5) y que dicha Asociación había negociado la suma de 4000 dólares y dos pasajes de barco en la esperanza de que el científico aceptase llevar a cabo un *tour* de conferencias por la Argentina. La nota de la Asociación también señalaba el interés del científico de que fuese una institución académica la que lo invitase, y por ese motivo, sabiendo del interés de la UBA, la Asociación consideraba apropiado que fuese esa casa de altos estudios la encargada de cursar la invitación.

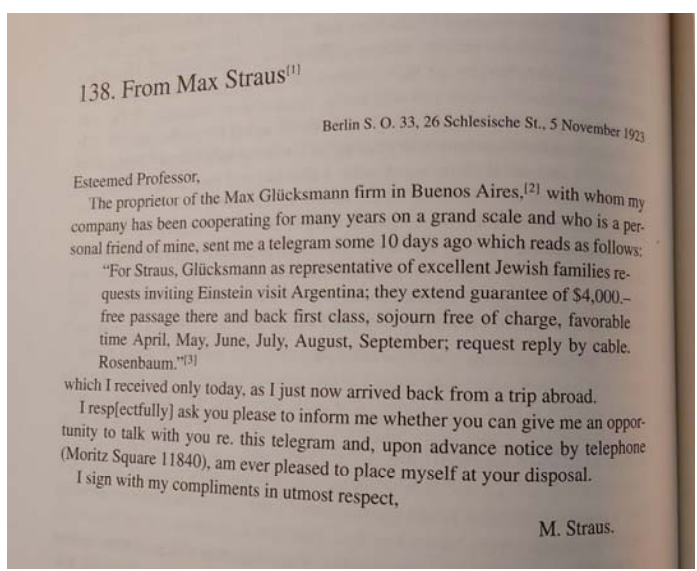

**Figura 5.** Traducción al inglés de la carta que Max Straus, director de la compañía discográfica Carl Lindström A.G., le enviara a Einstein. Con fecha del 5 noviembre de 1923, hace referencia a un telegrama que había enviado diez días antes desde Buenos Aires Max Glücksmann, "representante de familias judías de excelencia", con la invitación y oferta de honorarios, viáticos y alojamiento para la posible visita del científico a la Argentina (Einstein, 2015: Documento 138).

Finalmente, para ayudar con la logística y luego de las negociaciones locales, la Asociación Hebraica contribuiría con 4660 pesos (el equivalente de unos 1500 dólares de entonces) para afrontar parte de los honorarios del visitante. Por su parte, el Consejo Superior de la UBA aprobó emplear 2500 dólares para financiar la visita de Einstein, suma que, junto a la contribución más reducida de la Asociación Hebraica, llevaría el total a los 4000 dólares que el científico había considerado apropiado para materializar su visita al país. La UBA también se comprometía a cubrir la mitad del costo de los dos pasajes de barco en primera clase, en caso de que las tratativas de conseguir fondos del gobierno argentino para ese fin no dieran sus frutos (en un principio, Einstein iba a viajar acompañado, no por su esposa sino por Margot Einstein, su hijastra, algo que al final no sucedió). Días más tarde, en enero de 1924, la Institución Cultural Argentino-Germana también aportó una pequeña suma de dinero (una donación de 1500 pesos) para contribuir a financiar la visita. Recordemos que la comunidad alemana en Argentina había quedado algo dolida por la actitud pacifista de Einstein y su oposición a la escalada militarista durante la Primera Guerra mundial y, quizás peor, debido a su renuncia a la ciudadanía alemana en su juventud. Hay registros que certifican su inicial resistencia a avalar la visita de Einstein a nuestro país (Ortiz, 1995).

Por otra parte, una carta escrita en francés por los máximos representantes de la Asociación Hebraica de Buenos Aires y fechada el 9 de enero de 1924, le aclaraba a Einstein todos los puntos en los que el científico podía aún guardar alguna duda sobre su eventual visita a la Argentina. Los firmantes de la misiva eran Mauricio Nirenstein, el ya mencionado secretario de la UBA y, a su vez, también secretario de la Asociación Hebraica, e Isaac Starkmeth, el presidente de la Asociación (Dujovne, 2015). Algunos aspectos interesantes de la carta de la Asociación Hebraica son, por ejemplo, que deja en claro que Max Straus, quien envió la carta a Einstein en representación de Glücksmann, a su vez "representante de familias judías de excelencia" [Fig. 5], no se reunió con el científico sino con Elsa, llamada "Madame Einstein" en la carta, quien era ya conocida por su rol de "filtro" ante quienes intentaban aproximarse, de la manera que fuera, a su marido. En ausencia del sabio, durante ese par de meses convulsionados de fin de 1923, Elsa se había encargado de transmitir al emisario todos los requerimientos que debían satisfacerse para que el viaje sudamericano pudiese llevarse a cabo.

Es por ello por lo que los representantes de la Hebraica respondían, punto por punto y de manera muy concisa, a todos los interrogantes de los Einstein (Einstein, 2015: Documento 193). Así, dejaban en claro que era la UBA la institución académica que hacía la invitación, y adjuntaban copias de diarios locales que detallaban las resoluciones del Consejo Superior y demás noticias que avalaban esa invitación; que otras cuatro universidades argentinas se unían a la UBA para recibirlo (las universidades de Córdoba, de La Plata, del Litoral y de Tucumán), y que dispondrían de fondos para hacerlo; que se habían comenzado las tratativas para que también la Universidad de Montevideo, en Uruguay, lo recibiera, la que dispondría de 1000 pesos oro adicionales; además, le ofrecían a Einstein la posibilidad de que también visitara la Universidad de Santiago, en Chile, en cuyo caso le sugerían que el regreso a Europa podría hacerse por el Océano Pacífico en lugar de regresar por el Atlántico. No olvidaban, por supuesto, aclarar los montos detallados de los honorarios del visitante y le aseguraban que de ninguna manera su presencia en el país sería aprovechada para cualquier tipo de "publicidad o de propaganda de la naturaleza que fuese", puesto que "todas las cuestiones de carácter político o nacionalista están excluidas del programa de la Asociación Hebraica" (esta última frase estaba subrayada con lápiz rojo). Le aseguraban, además, que la Hebraica no decidía sobre su programa académico en los países a visitar; solo pedían poder "ofrecerle una recepción y escuchar [de él] una o dos conferencias populares sobre el tema que juzgará más apropiado al carácter de nuestra institución". Por último, le señalaban que "algunos israelitas notables" ya habían ofrecido sus casas para que el científico pernoctara durante sus días en Buenos Aires y esperaban poder ofrecerle todas las comodidades "para trabajar y reposarse de manera satisfactoria" durante su estadía en la ciudad. Esta carta fue recibida por Einstein una vez de regreso en Berlín y respondida el 8 de marzo del mismo año [Fig. 6].

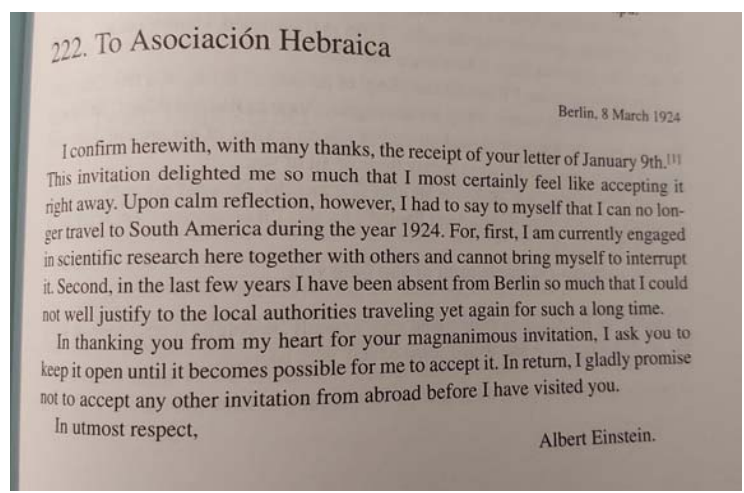

**Figura 6.** Traducción al inglés de la respuesta de Einstein del 8 de marzo de 1924 a la carta de la Asociación Hebraica con la invitación para visitar la Argentina. El científico agradece con diplomacia, pero señala que le será imposible desplazarse durante ese

año por diversos motivos. Promete, sin embargo, que no aceptará ninguna otra invitación antes de cumplir con la visita a la Argentina (Einstein, 2015: Documento 222).

A partir de ese momento, el Consejo Superior de la UBA recibió noticias frecuentes del estado de las negociaciones con el futuro visitante. En mayo de 1924, la Universidad fue informada de la correspondencia intercambiada por Einstein con el representante del Ministerio de Relaciones Exteriores de la Argentina en Berlín y con la Asociación Hebraica. La visita se fijó para fines de marzo del año siguiente, es decir, hacia el comienzo del año académico de 1925. Finalmente, y luego de un viaje de tres semanas en barco, Einstein tocó suelo argentino en la mañana del 25 de marzo de 1925 (El Hogar, 1925) y, quizá como él mismo podría haberlo anticipado, se encontró frente a una abultada agenda de actividades -algunas científicas, sí, pero por sobre todas las cosas, muchas actividades protocolares y sociales- que le dejaron muy poco tiempo para trabajar y pensar, o siquiera para descansar.

**V. Palabras finales**

Como vimos, las publicaciones sobre temas relacionados con la física y la matemática de la relatividad comienzan a aparecer en la Argentina en los primeros años del siglo XX. Luego de los resultados del eclipse de Sol de 1919 la relatividad fue el tema científico que quizás más atrajo la atención en los círculos académicos internacionales. El impacto de este resurgimiento del interés relativista en nuestro país, en gran medida, se lo debemos a Leopoldo Lugones, notorio poeta y científico aficionado, primero con su conferencia de 1920 y luego con su "activismo" a favor de traer a Einstein a la Argentina. La posibilidad real de una visita del científico a nuestro país fue concebida a mediados de 1922, en coincidencia con los días en los que Einstein comenzaba a experimentar las dificultades de la vida en Alemania. Para entonces, y en el par de años siguientes, la producción de trabajos y monografías sobre temas de física y filosofía relacionados con la nueva imagen del espacio-tiempo había aumentado considerablemente. Todos estos factores contribuyeron fuertemente a hacer de la visita de Einstein en 1925 un verdadero y memorable evento cultural.

No deja de sorprender, aún hoy, que para que se explicaran los fundamentos y los avances más recientes de la teoría de la relatividad en Sudamérica, la intelectualidad argentina de esos años buscó, precisamente, la palabra del creador de la teoría. En contraste con esto, sabemos que sólo unos pocos años más tarde llegó a Buenos Aires otro relativista eminente, Paul Langevin quien, con el auspicio del Instituto Francés de la Universidad de París en Buenos Aires visitó el país con un costo sumamente reducido para los cofres de la universidad; además, Langevin rechazó los honorarios especiales que le ofreció la universidad por sus conferencias. Por supuesto, desde el inicio de la Primera Guerra Mundial, y sobre todo desde que Einstein firmara a fines de 1914 el *Manifiesto a los europeos*, el llamado "contra-manifiesto" que se oponía en Alemania a la declaración de la Guerra, su nombre paulatinamente crecía entre los máximos exponentes pacifistas. Esto, junto a la adhesión a las nuevas ideas de cambio que dominaban en una parte de la intelectualidad de Buenos Aires de ese momento, muchas surgidas a partir de la Reforma Universitaria de 1918, hacía que la figura y la obra de Einstein fueran muy representativas de varios círculos académicos locales.

La visita de Einstein a la Argentina, el último de una serie de largos viajes internacionales del científico, de más está decir, tiene también una dimensión importante para el estudio del desarrollo de la comunidad judía local y de sus relaciones con el movimiento intelectual contemporáneo, el que es discutido en detalle, por ejemplo, en (Ortiz, 1995; Dujovne, 2015). Como vimos, unos años después del eclipse de 1919 que lo volviera una figura mundial y ya en momentos en que la situación social de Berlín se volvía peligrosa para el célebre científico, miembros de la creciente comunidad judía de la Argentina unieron sus fuerzas y recursos con otras instituciones e individuos -universidades, asociaciones y miembros de la cultura porteña- de tal manera de hacer posible la visita de Einstein en el corto plazo. Ya desde el año 1923 la Asociación Hebraica había tomado la figura del eminente científico judío y pacifista como una excelente oportunidad para iniciar su programa cultural. Este evento, de materializarse exitosamente, le permitía a la vez mostrar a la comunidad intelectual de la Argentina sus firmes lazos con Europa y su gran capacidad para atraer destacadas personalidades a sus tierras. Mostraba, además, su influencia y el veloz progreso que, en tan solo unas décadas, había logrado la comunidad judía -con sus miles de inmigrantes desde fines del siglo XIX- que se había firmemente afincado en los países sudamericanos.

Las numerosas actividades realizadas por el científico en su apretada agenda argentina, y también, aunque en menor medida, en los dos países vecinos, Uruguay y Brasil, fueron ya detalladas en varias publicaciones anteriores (Ortiz, 1995; Ortiz y Otero, 2001; Gangui y Ortiz, 2005; Asúa y Hurtado de Mendoza, 2006; Gangui y Ortiz, 2008; Einstein, 2008). En este trabajo nos hemos limitado a poner el foco en los importantes antecedentes y en los diversos intereses convergentes tanto de grupos argentinos como del mismo Einstein que, pocos años antes de la visita, posibilitaron su viaje a Sudamérica. Así, hemos podido repasar una pequeña parte de la complicada y costosa logística desplegada en Buenos Aires para acercar al padre de la relatividad a sus orillas.